\input amstex
\documentstyle{amsppt}
\magnification=\magstep1 \baselineskip=18pt \hsize=6truein
\vsize=8truein

\topmatter
\title A positive mass theorem on asymptotically hyperbolic manifolds with corners along a hypersurface \endtitle
\author  Vincent Bonini and Jie Qing \endauthor

\abstract In this paper we take an approach similar to that in [M]
to establish a positive mass theorem for spin asymptotically
hyperbolic manifolds admitting corners along a hypersurface. The
main analysis uses an integral representation of a solution to a
perturbed eigenfunction equation to obtain an asymptotic expansion
of the solution in the right order. This allows us to understand the
change of the mass aspect of a conformal change of asymptotically
hyperbolic metrics.
\endabstract

\subjclass{Primary 53C21; Secondary 53C24, 83C40}\endsubjclass

\keywords{Asymptotically hyperbolic manifold, positive mass theorem,
decay rate, conformal deformation, integral
representation}\endkeywords

\thanks The first named author supported by MSRI Postdoctoral Fellowship.\endthanks
\thanks The second named author supported partially by NSF grant DMS 0402294 \endthanks

\leftheadtext{positive mass theorem} \rightheadtext{Bonini and Qing}

\address
Vincent Bonini, Mathematical Science Research Institute, Berkeley,
CA 94720
\endaddress

\email vbonini\@msri.org \endemail

\address
Jie Qing, Department of Mathematics, UC, Santa Cruz, CA 95064
\endaddress

\email qing\@math.ucsc.edu \endemail
\endtopmatter

\document

\head 1. Introduction\endhead

In this paper we study the change of mass aspect for asymptotically
hyperbolic manifolds under a conformal change of metric and
establish a positive mass theorem for a class of asymptotically
hyperbolic manifolds admitting corners along a hypersurface. This
work follows an approach similar to that in [M]. The dimensions of
all manifolds concerned in this paper are greater than 2. Positive
mass theorems for asymptotically hyperbolic manifolds have been
studied in many works, notably in [Mo] [AD] [W] [CH]. A Riemannian
manifold $(M, \ g)$ with corners along a hypersurface $\Sigma$ is a
manifold that is separated by an embedded hypersurface
$\Sigma\subset M$ such that each individual part is a smooth
Riemannian manifold and the metric $g$ is continuous across the
hypersurface $\Sigma$. An asymptotically hyperbolic manifold with
corners along a hypersurface is a Riemannian manifold with corners
along a hypersurface with one part compact and the other part
asymptotically hyperbolic. The issue at hand is to investigate the
validity of a positive mass theorem for asymptotically hyperbolic
manifolds with corners along a hypersurface if each part satisfies
the scalar curvature condition. A good motivation given in [M] to
initiate the study of such question is to use the Ricatti equation
$$
R = R_\Sigma  - (|A|^2 + H^2) - 2 \frac {\partial H}{\partial n},
\tag 1.1
$$
which allows one to consider the scalar curvature in distributional
sense across the hypersurface. It also turns out to relate to a
notion of quasi-local mass in relativity (cf. [B] [M] [ST1] [ST2]).
It is desirable to have a non-negative quantity associated with a
compact domain $\Omega$ of an asymptotically hyperbolic manifold
$M$, which is zero if and only if $\Omega$ can be isometrically
embedded into the hyperbolic space and converges to the total mass
when $\Omega$ exhausts $M$. Analogous to the suggestion for the
asymptotically flat setting in [B], a natural candidate for such a
quantity is given by taking the infimum of the total mass over the
class of all asymptotically hyperbolic manifolds in which $\Omega$
can be isometrically embedded and to which positive mass theorem can
apply. For more details readers are referred to [B] [M] [ST1] [ST2].

In case of an asymptotically hyperbolic manifold with corners along
a hypersurface we will call the compact part the inside and the
non-compact part the outside. We will denote the mean curvature of
the hypersurface with respect to the inside metric in the outgoing
direction by $H_-$ and the mean curvature of the hypersurface with
respective to the outside metric in the direction inward to the
outside by $H_+$. Our main theorem is as follows:

\proclaim{Theorem 1.1} Suppose that $(M^n, \ g)$ is a spin
asymptotically hyperbolic manifold of dimension $n\geq 3$ with
corners along a hypersurface. And suppose that the scalar curvature
of both the inside and outside metrics are greater than or equal to
$-n(n-1)$ and that
$$
H_- (x) \geq H_+ (x)
$$
for each $x$ on the hypersurface. Then, if in a coordinate system at
the infinity,
$$
g = \sinh^{-2}\rho (d\rho^2 + g_0 + \frac {\rho^n}n h +
O(\rho^{n+1})),
$$
then
$$
\int_{S^{n-1}} \text{Tr}_{g_0}h (x)dvol_{g_0}(x) \geq
|\int_{S^{n-1}}x\text{Tr}_{g_0}h (x)dvol_{g_0}(x)|. \tag 1.2
$$
\endproclaim

In [W] the vanishing of the mass is proved to imply the
asymptotically hyperbolic manifold is isometric to the hyperbolic
space. However, we did not find it is a straightforward consequence
to have the same conclusion in our context nor did Miao in [M] in
the context of asymptotically flat manifolds. We will give an
affirmative answer to this question in a forthcoming paper. We would
like to point out though it is easy to see that the scalar curvature
should be the constant as the hyperbolic space.

We adopt an approach from [M] to smooth the corners, then
conformally deform the metric so that the scalar curvature is
greater than or equal to $-n(n-1)$ and then apply the positive mass
theorem in [W]. Instead of solving an equation which is a
perturbation of Laplace equation as in [M] [SY] for asymptotically
flat case, we realize, with our experience in [Q] [BMQ], that we
should consider an equation which is a perturbation of the
eigenfunction equation
$$
-\Delta v + nv = 0 \tag 1.3
$$
on an asymptotically hyperbolic manifold, where
$$
\Delta = \sum_{i=1}^n \frac{\partial^2}{\partial x_i^2}
$$
on $R^n$ in our notation in this paper. We also learned that in fact
in each case the operator is simply the linearization of the Yamabe
equation at the constant scalar curvature one. One of the
consequences of this consideration gives hope that $v$ decays in the
right order to allow us to estimate the change of mass aspect after
a conformal change of metric while another is the following key
observation.

\proclaim{Lemma 1.2} Suppose that $(M^n, \ g)$ is a Riemannian
manifold and $v$ is a positive smooth solution to the linear
equation
$$
-\Delta v + nv - \frac {n-2}{4(n-1)}(R+n(n-1))^-v = \frac
{n-2}{4(n-1)}(R+n(n-1))^-. \tag 1.4
$$
Then the scalar curvature of the metric $g_v = (1+v)^\frac 4{n-2}g$
satisfies
$$
R_{g_v} \geq - n(n-1). \tag 1.5
$$
\endproclaim

To find a solution $v$ to (1.4) we use the analysis of weighted
function spaces and uniformly degenerate elliptic equations, which
are well developed in, for example, [A] [AC] [GL] [L1] [L2] [Mz]
[MM]. The positivity of the solution $v$ to (1.4) follows from a
clever use of a generalized maximum principle in [PW]. We have
noticed that the existence of the expansion of the solution $v$ was
studied in [MM] [AC]. But we need the explicit formula to estimate
the change of mass aspects here. We followed the approach taken in
[SY] which used an integral representation to obtain an asymptotic
expansion. To obtain an integral representation we used an explicit
formula for the fundamental solution to the eigenfunction equation
in the hyperbolic space
$$
G_H(x, y) = \frac {c_n}{\sinh^{n-2}d_H(x, y)\cosh^2d_H(x,
y)}\theta(\cosh d_H(x, y)), \tag 1.7
$$
where $d_H(x, y)$ is the hyperbolic distance between $x$ and $y$ in
hyperbolic space $H^n$,
$$
c_n = \frac 1{(n-2)\text{vol}(S^{n-1})},
$$
$$
\theta (s) = \frac 1{\theta_0}(1 + \sum_{i=2}^\infty\prod_{j=2}^i(1
- \frac {n}{2j + n -1})s^{-2i+2}) \tag 1.8
$$
and
$$
\theta_0 = 1 + \sum_{i=2}^\infty \prod_{j=2}^i(1 - \frac {n}{2j + n
-1}). \tag 1.9
$$
For more detailed account on the above generalized eigenfunctions
please see [AC] [MM]. Thus

\proclaim{Lemma 1.3} Suppose that $(M^n, \ g)$ is an asymptotically
hyperbolic manifold, $M_c$ is a compact set in $M$ and $r_0$ is a
large number. Let
$$
x = \psi(p): M\setminus M_c \to R^n\setminus B_{r_0}(0),
$$
be a coordinate at the infinity in which
$$
g = \sinh^{-2}\rho (d\rho^2 + g_0 + \frac {\rho^n}n h +
O(\rho^{n+1})),
$$
where $\sinh\rho = |x|^{-1}$. Suppose that $v\in C^{2,
\alpha}_\delta(M)$ with $\delta > 0$ solves the equation
$$
-\Delta v + nv + fv = w,
$$
with
$$
f\in C^{0, \alpha}_\kappa(M) \ \text{and} \ w \in C^{2,
\alpha}_\eta(M),
$$
for some $\kappa > 2$ and $\eta>n+1$. Then
$$
v(x) = A(\frac x{|x|})|x|^{-n} + O(|x|^{-(n+1)}) \tag 1.12
$$
for some function $A$ on $S^{n-1}$.
\endproclaim

Note that the function $A(\frac {x}{|x|})$ in the above lemma in our
proof will be given as a sum of several integrals which later allow
us to estimate the size of change of the mass aspects, please see
Lemma 6.5 in this note.

The paper is organized as follows: Section 2 is devoted to
establishing an isomorphism theorem for a class of uniformly
degenerate operators based on work in [L2]. In Section 3 we
introduce a linear equation whose solution gives a conformal factor
for a metric with the scalar curvature greater than or equal to
$-n(n-1)$. In Section 4 we derive an explicit formula for the
fundamental solutions to the eigenfunction equation on hyperbolic
space $H^n$. In Section 5 we use the standard fundamental solution
to construct an approximate fundamental solution on an
asymptotically hyperbolic manifold. This gives us an integral
representation of a solution to the eigenfunction equation and the
desired asymptotic expansion. In Section 6 we prove our main theorem
by calculating the mass aspect of the deformed metric and applying
the positive mass theorem in [W].

\vskip 0.1in\noindent{\bf Acknowledgement} \quad We would like to
thank the referee for a very thorough reading of our manuscript and
many constructive suggestions.

\head 2. Analytic preliminaries\endhead

In this section we discuss some preliminaries of the analysis on
weakly asymptotically hyperbolic manifolds. Let $\bar M^n$ be a
smooth compact $n$-dimensional manifold with boundary $\partial M$
and $M^n$ be its interior. A nonnegative smooth function $\rho$ on
$\bar M$ is said to be a defining function for $\partial M$ if
$$
\aligned \rho > 0 & \quad \text{in $M$}\\ \rho = 0 & \quad \text{on
$\partial M$}\endaligned
$$
and $d\rho$ never vanishes on $\partial M$. For any non-negative
integer $m$ and any $0 \leq \beta < 1$, a smooth Riemannian metric
$g$ on $M$ is then said to be conformally compact of class $C^{m,
\beta}$ if for any defining function $\rho$ for $\partial M$, the
conformal metric $\bar {g} = \rho^2 g$ extends as a $C^{m, \beta}$
metric on $\bar M$. The metric $\bar g$ restricted to $T (\partial
M)$ induces a metric $\hat {g}:= \bar {g} |_{T (\partial M)}$ on
$\partial M$ which rescales upon change in defining function and
therefore defines a conformal structure $[\hat {g}]$ on $\partial M$
called  the conformal infinity of $(M, g)$.

When $m + \beta \geq 2$, a straightforward computation as in [Mz]
shows that the sectional curvatures of $g$ approach $-|d
\rho|^2_{\bar {g}}$ at $\partial M$. As in [BMQ], we define weakly
asymptotically hyperbolic manifolds as follows:

\proclaim{Definition 2.1} A connected complete Riemannian manifold
$(M^{n}, \ g)$ is said to be weakly asymptotically hyperbolic of
class $C^{m, \beta}$ if $g$ is conformally compact of class $C^{m,
\beta}$ with $m + \beta \geq 2$ and $|d \rho|^2_{\bar {g}}=1$ on
$\partial M$ for a defining function $\rho$.
\endproclaim

We will use the definitions of weighted function spaces from the
papers of Lee [L1] [L2](see also [GL] [A]). Let $(M^n, \ g)$ be a
weakly asymptotically hyperbolic manifold and let $\rho$ be a
defining function. The weighted H\"{o}lder spaces are defined, for
$\delta\in R$,
$$
C^{k, \alpha}_\delta(M) : =\rho^\delta C^{k, \alpha}(M) =
\{\rho^\delta u: u\in C^{k, \alpha}(M)\} \tag 2.1
$$
with the norm
$$
\|u\|_{C^{k, \alpha}_\delta(M)} : = \|\rho^{-\delta}u\|_{C^{k,
\alpha}(M)}.
$$
The weighted Sobolev spaces are defined, for $\delta\in R$,
$$
W^{k, p}_{\delta} (M) : = \rho^\delta W^{k, p}(M) = \{\rho^\delta u:
u\in W^{k, p}(M)\} \tag 2.2
$$
with the norm
$$
\|u \|_{W^{k, p}_\delta} : = \|\rho^{-\delta} u\|_{W^{k, p}(M)}.
$$
We recall the following weighted Sobolev embedding theorem from
[L2].

\proclaim{Lemma (Sobolev Embedding)} Let $(M^{n}, g)$ be weakly
asymptotically hyperbolic manifold of class $C^{m, \beta}$ and $U
\subset M$ an open subset. For $1 < p, q < \infty$, $0 < \alpha <
1$, $\delta \in R$, $1 \leq k \leq m$, and $k+\alpha \leq m+\beta$,
the inclusions
$$
W^{k, q}_{\delta}(U) \hookrightarrow  W^{j, p}_{\delta}(U) \quad
\text{for} \quad k - \frac{n}{q} \geq j - \frac{n}{p} \tag 2.3
$$
and
$$
W^{k, p}_{\delta}(U) \hookrightarrow C^{j, \alpha}_{\delta}(U) \quad
\text{for} \quad k - \frac{n}{p} \geq j + \alpha \tag 2.4
$$
are continuous.
\endproclaim

The readers are referred to [L2] (see also [GL] [A] [L1]) for a more
complete discussion of properties of the weighted H\"{o}lder and
Sobolev spaces on weakly asymptotically hyperbolic manifolds. Our
goal in this section is to derive an isomorphism result from [GL]
[L2], particularly Theorem C in [L2], for the operator $-\Delta + n
+ f$. We first state a simpler version of Theorem C in [L2].

\proclaim{Lemma 2.2} Suppose that $(M^n, \ g)$ is a weakly
asymptotically hyperbolic manifold of class $C^{m, \beta}$. Let
$k+1+\alpha \leq m+\beta$ and $f\in C^{0, \alpha}_{\gamma}$ for some
$\gamma >0$. Then
$$
-\Delta + n + f: C^{2, \alpha}_\delta (M)\to C^{0, \alpha}_\delta(M)
$$
is a zero index Fredholm operator whenever $\delta \in (0, n)$. The
possible kernel is the $L^2$-kernel of $-\Delta + n + f$.
\endproclaim

Then we derive an isomorphism result by asking that $-\Delta + n +
f$ is a perturbation of $-\Delta + n$ with the negative part of $f$
small in integral sense. We will denote
$$
f = f^+ - f^-
$$
where $f^+ = \max\{f, 0\}$ and $f^- = - \min\{f, 0\}$.

\proclaim{Proposition 2.3}Suppose that $(M^n, \ g)$ is a weakly
asymptotically hyperbolic manifold of class $C^{m, \beta}$. Let $4
\leq m+\beta$ and $f\in C^{0, \alpha}_{\gamma}$ for some $\gamma
> 0$. Then there is a positive number $\epsilon_0$ such that, if
$$
(\int_M |f^-|^\frac n2dvol)^\frac 2n \leq \epsilon_0, \tag 2.5
$$
then
$$
-\Delta + n + f: C^{2, \alpha}_\delta (M)\to C^{0, \alpha}_\delta(M)
\tag 2.6
$$
is an isomorphism when $\delta\in (0, n)$.
\endproclaim

\demo{Proof} Suppose that $v$ is a function in the $L^2$-kernel of
the operator $-\Delta + n + f$. Due to some standard weighted $L^2$
estimates (cf. Lemma 4.8 in [L2], for instance) we know that $v\in
W^{2, 2}(M)$ and solves the equation
$$
-\Delta v + nv + fv = 0. \tag 2.7
$$
Let $\rho$ be a geodesic defining function for the weakly
asymptotically hyperbolic manifold $(M^n, \ g)$. For $\epsilon > 0$
let
$$
M_{\epsilon} = \{p\in M: 0 < \rho(p) < \epsilon \}.
$$
Multiplying (1) by $v$ and integrating by parts over $M \backslash
M_{\epsilon}$ we see
$$
\aligned
0 &= \int_{M \backslash M_{\epsilon}} {-v \Delta v + f v^{2} + n v^{2}}  \\
&= \int_{M \backslash M_{\epsilon}}(|\nabla v |^{2} + n v^{2}) +
\int_{M \backslash M_{\epsilon}}f v^{2} + \int_{\{ \rho = \epsilon
\}}{ v \frac{\partial v}{\partial \vec{n}} \ d\sigma}.
\endaligned
$$
Now $v \in W^{2,2}(M)$ so for a fixed small number $\epsilon_1>0$
$$
\int_0^{\epsilon_1} \int_{\rho = s}|v||\nabla v|d\sigma\frac {ds}s =
\int_{M\setminus M_{\epsilon_1}}|v||\nabla v|  < \infty.
$$
Therefore, there is a sequence of $\epsilon_i\to 0$ such that
$$
\int_{\rho= \epsilon_i} |v||\nabla v|d\sigma \to 0,
$$
which implies
$$
\int_M (|\nabla v|^2 + nv^2) = - \int_M fv^2.
$$
Then, by H\"{o}lder inequality,
$$
\int_M (|\nabla v|^2 + nv^2) \leq \int_M f^-v^2 \leq (\int_M
(f^-)^\frac n2)^\frac 2n (\int_M v^\frac {2n}{n-2})^{1 -\frac 2n}.
$$
Next we apply the Sobolev embedding Theorem and obtain
$$
\int_M (|\nabla v|^2 + nv^2) \leq C(\int_M (f^-)^\frac n2)^\frac 2n
\int_M(|\nabla v|^2 + v^2), \tag 2.8
$$
where $C$ here is the Sobolev constant, which is independent of $v$.
Thus, for
$$
\epsilon_0 = \frac 1{2C},
$$
we may conclude that $v=0$. So the proposition follows from Lemma
2.2.
\enddemo

\head 3. Conformal deformations \endhead

In this section we discuss the conformal deformation of the scalar
curvature on an asymptotically hyperbolic manifold $(M^n, \ g)$.
This idea comes from the work in [SY] where the analogous situation
was treated in the context of asymptotically flat manifolds.

\proclaim{Lemma 3.1} Suppose that $v$ is a positive solution to the
following equation
$$
-\Delta v + n v - \frac {n-2}{4(n-1)}(R+n(n-1))^-v = \frac
{n-2}{4(n-1)}(R+n(n-1))^- \tag 3.1
$$
on a manifold $(M^n, \ g)$. Then
$$
R[(1+v)^\frac 4{n-2}g] \geq -n(n-1).
$$
\endproclaim

\demo{Proof} Let $u = 1+ v$. Then
$$
\aligned -\Delta u & + \frac {n-2}{4(n-1)}Ru = -\Delta v + \frac
{n-2}{4(n-1)}(R+n(n-1))u - \frac{n(n-2)}4 u \\ & \geq -\Delta v + nv
- \frac{n-2}{4(n-1)}(R+n(n-1))^-v -
\frac{n-2}{4(n-1)}(R+n(n-1))^-\\& \quad\quad - nv -
\frac{n(n-2)}4(1+v)\\ & = - \frac {(n-2)}{4(n-1)} n(n-1) \frac {1 +
\frac 4{n-2}\frac v{1+v}}{(1+v)^{\frac 4{n-2}}}u^{\frac{n+2}{n-2}}.
\endaligned
$$
Hence to prove the lemma is to show that
$$
1 + \frac 4{n-2} - \frac 4{n-2}\frac 1{1+v} \leq (1+v)^{\frac
4{n-2}}. \tag 3.2
$$
We differentiate the two sides with respect to $v$ and compare
$$
\frac 4{n-2} (1+v)^{-2} < \frac 4{n-2} (1+v)^{\frac 4{n-2} - 1}.
$$
Therefore, by the fact that the two sides are the same when $v=0$,
the lemma follows.
\enddemo

The rest of this section is devoted to solving for a positive
solution to the equation
$$
(-\Delta + n + f)v = h \tag 3.3
$$
on an asymptotically hyperbolic manifold $(M^n, \ g)$ with the
function $f$ suitably small in an integral sense. By the isomorphism
proposition in the previous section we know, for $\delta\in (0, n)$
and each $h\in C^{0, \alpha}_\delta(M)$, there is a unique solution
$v\in C^{2, \alpha}_\delta(M)$ to the equation (3.3). Hence what
really need to do is to show that $v>0$ in $M$. For simplicity we
will denote
$$
f = -\frac {n-2}{4(n-1)}(R+n(n-1))^- \leq 0.
$$

\proclaim{Proposition 3.2} Suppose that $(M^n, \ g)$ is a weakly
asymptotically hyperbolic manifold of class $C^{m, \beta}$ with
$m+\beta \geq 4$. Let $\epsilon_0$ be the small positive number in
Proposition 2.3 in the previous section and $\alpha \in (0, 1)$.
Suppose that $f\in C^{0, \alpha}_{\delta}(M)$ for some $\delta\in
(0, n)$ and that
$$
(\int_M |f|^\frac n2)^\frac 2n \leq \epsilon_0. \tag 3.4
$$
Then there is a positive solution $v\in C^{2, \alpha}_\delta(M)$ to
the equation
$$
-\Delta v + nv + fv = -f. \tag 3.5
$$
\endproclaim

\demo{Proof} We first prove that $v$ has to be nonnegative in $M$.
Assume otherwise that $v$ is negative somewhere in $M$ so that
$$
v_-= \min\{v(p): p\in M\} < 0.
$$
Let us consider instead the function $u = v+v_0$ for a small
positive number $v_0< \min\{1, -\frac {v_-}2\}$. Then
$$
-\Delta u + nu + f u = -f(1 - v_0) + nv_0 > 0
$$
in $M$ and $\min\{u(p): p\in M\} < 0$. Since $v\in C^{2,
\alpha}_\delta(M)$ for $\delta > 0$, for a geodesic defining
function $\rho$, we may assume that
$$
u > 0 \quad \text{on $\partial(M\setminus M_\tau) = \{p\in M: \rho
(p) = \tau\}$}
$$
provided that $\tau > 0$ is sufficiently small. Now we are going to
apply the generalized maximum principle in Section 2.5 in [PW] to
the function $u$ on the manifold $M\setminus M_\tau$. According the
generalized maximum principle what we need is to verify that the
first eigenvalue of the operator $-\Delta + n + f$ on the domain
$M\setminus M_{\tau'}$ for some $\tau' < \tau$ with Dirichlet
boundary condition is positive. Therefore, for any $\phi\in
C^\infty_c(M\setminus M_{\tau'})$, we consider the ratio
$$
\aligned & \frac {\int_M (|\nabla\phi|^2 + n \phi^2 +
f\phi^2)}{\int_M \phi^2} \\ & \geq\frac 1{\int_M\phi^2} (\int_M
(|\nabla\phi|^2 + \phi^2) - C(\int_M |f|^\frac n2)^\frac
n2(\int_M(|\nabla\phi|^2+\phi^2))) \\& \geq \frac 12. \endaligned
$$
Thus the first eigenvalue of the operator $-\Delta + n + f$ on the
domain $M\setminus M_{\tau'}$ with the Dirichlet boundary condition
is always positive. We may apply Theorem 10 in Section 2.5 of the
book [PW] to the function $\frac u\phi$, where $\phi$ is the
positive first eigenfunction over $M\setminus M_{\tau'}$, to obtain
a contradiction. Therefore $v$ is nonnegative in $M$. To show that
$v$ is in fact positive in $M$, for each $\tau > 0$, we apply the
Hopf strong maximum principle to the function $\frac v{\phi}$ on the
domain $M\setminus M_\tau$, where $\phi$ is the positive first
eigenfunction over $M\setminus M_{\tau'}$ for any $0< \tau' < \tau$.
Thus the proof is complete.
\enddemo

\head 4. The fundamental solutions on the hyperbolic space\endhead

The materials in this section are well known and readers are
refereed to [A] [AC] [L2] [MM] for more detailed account on the
references. But for the convenience of the readers we will present a
construction briefly. Let us first recall the definition of the
hyperbolic space as a hyperboloid in the Minkowski space-time. The
Minkowski space-time is $R^{n+1}$ equipped with the Minkowski metric
$- dt^2 + |dx|^2$ for $(t, x)\in R^{n+1}$. The upper hyperboloid is
the submanifold
$$
H^n = \{(t, x)\in R^{n+1}: - t^2 + |x|^2 = -1,  \ t
> 0\}. \tag 4.1
$$
Hence
$$
(H^n, \ g_H) = (R^n, \ \frac {(d|x|)^2}{1 + |x|^2} + |x|^2
g_{S^{n-1}}), \tag 4.2
$$
where $g_{S^{n-1}}$ is the standard metric on the unit round
$(n-1)$-sphere. We want to find the solution to the equation
$$
-\Delta_{H^n} G_0(x) + n G_0(x) = \delta_0(x), \tag 4.3
$$
which defines the Green's function in $x$ centered at the origin of
the differential operator $-\Delta + n$ on hyperbolic space $H^n$.
We first compute, for $r = |x|$,
$$
(-\Delta_{H^n} + n) r^{-n+2} t^{-k} = - (k-2)(k + n
-1)r^{-n+2}t^{-k} + k(k+1) r^{-n+2} t^{-k-2}.
$$
We then observe inductively that, for even number $k$,
$$
\aligned (-\Delta_{H^n} + n)&(r^{-n+2}(t^{-2} + \frac {2\cdot
3}{2(n+3)}t^{-4} + \cdots + \frac {2\cdot 3\cdot 4\cdot 5\cdots (k-1)}{(2(n+3)\cdots (k-2)(k+ n -1)}t^{-k})\\
& = \frac {2\cdot 3\cdot 4\cdot 5\cdots  k\cdot (k+1)}{(2(n+3)\cdots
(k-2)(k + n- 1)}r^{-n+2}t^{-k-2}\endaligned
$$
Therefore we consider the function
$$
\tilde \theta(t) = (1 + \sum_{i=2}^\infty \prod_{j=2}^i(1 - \frac
{n}{2j + n -1}) \frac 1{t^{2i-2}}). \tag 4.4
$$
Notice that the infinite series $\tilde\theta$ is obviously
convergent when $t>1$. In fact, when $t=1$, taking the logarithm of
the general term we see
$$
\log  \prod_{j=2}^i(1 - \frac {n}{2j + n -1}) \leq - \frac n2 \log
(i +\frac {n-1}2) + c(n)
$$
for some dimensional constant $c(n)$. Thus the infinite series
$$
\tilde\theta(1) = 1 + \sum_{i=2}^\infty \prod_{j=2}^i(1 - \frac
{n}{2j + n -1}) \tag 4.5
$$
converges for all $n\geq 3$. We set
$$
\theta (t) = \frac {\tilde\theta(t)}{\tilde\theta(1)} \tag 4.6
$$
and easily conclude that

\proclaim{Lemma 4.1} Let
$$
G_0(x) = \frac {\theta(t)}{(n-2)\text{vol}(S^{n-1})}\frac
1{r^{n-2}t^2}. \tag 4.7
$$
Then
$$
-\Delta_{H^n} G_0(x) + n G_0(x) = \delta_0(x)
$$
on hyperbolic space $H^n$.
\endproclaim

To write the fundamental solution at any point in the hyperbolic
space we want to express hyperbolic translation in the hyperboloid
model of hyperbolic space $H^n$. Recall that the changes of
coordinates between the ball model and hyperboloid model of the
hyperbolic space are
$$
x = \frac 2{1 - |\bar x|^2} \bar x, \quad t = \frac {1+|\bar x|^2}{1
- |\bar x|^2},
$$
and
$$
\bar x = \frac 1{1+t} x.
$$
Also recall that hyperbolic translation by $\bar b$ in the ball
model is given in [R] by
$$
\tau_{\bar b}(\bar x) = \frac {1 - |\bar b|^2}{|\bar x|^2|\bar b|^2
+2\bar x\cdot \bar b + 1}\bar x + \frac {|\bar x|^2 + 2 \bar x\cdot
\bar b + 1}{|\bar x|^2|\bar b|^2 +2\bar x\cdot \bar b + 1}\bar b,
\tag 4.8
$$
where $t_x = \sqrt{1+|x|^2}$ and $t_b = \sqrt{1+ |b|^2}$. Therefore
we have
$$
T_b(x) = x + t_x b + \frac {x\cdot b}{1 + t_b}b \tag 4.9
$$
with
$$
|T_b(x)| = \sinh d_H(x, \ -b). \tag 4.10
$$
One key fact here is that
$$
\cosh d_H(x, \ b) = t_x t_b - x\cdot b. \tag 4.11
$$
Thus
$$
G_H(x, y) = G_{y}(x) = G_0(T_{-y}(x)). \tag 4.12
$$
and explicitly
$$
G_H(x, y) = \frac {c_n}{\sinh^{n-2}d_H(x, y)\cosh^2d_H(x, y)} \theta
(\cosh d_H(x, y)) \tag 4.13
$$
where
$$
c_n = \frac 1{(n-2)\text{vol}(S^{n-1})}. \tag 4.14
$$

\head 5. Asymptotic behavior \endhead

So far, for a weakly asymptotically hyperbolic manifold $(M^n, \ g)$
with
$$
(R+n(n-1))^-\in C^{0, \alpha}_\delta \quad \text{and} \ \int_M
((R+n(n-1))^-)^\frac n2 \leq \epsilon_0^\frac n2,
$$
we have obtained a conformal deformation $g_v = (1+v)^\frac 4{n-2}
g$ such that
$$
R[g_v] \geq -n(n-1)
$$
and
$$
0 < v \in C^{2, \alpha}_\delta (M),
$$
provided that $\delta\in (0, n)$. Unfortunately the decay rate of
$v$ just misses the decay rate on which the mass aspect of an
asymptotically hyperbolic manifold is defined. We will use the
Green's function we constructed in the pervious section to obtain an
expansion at the infinity of the solution $v$ to the equation
$$
-\Delta v + nv - \frac {n-2}{4(n-1)}(R+n(n-1))^-v = \frac
{n-2}{4(n-1)}(R+n(n-1))^-.
$$
We follow the idea used in [SY] to write an integral representation
of the solution $v$ with the help of the approximate Green's
function $G_H(x, y)$ on the asymptotically hyperbolic manifold $M$.
Let us start with a definition of asymptotically hyperbolic
manifolds, which should be compared with the definition of weakly
asymptotically hyperbolic manifolds given in Section 2. Since we
will adopt the definition of mass aspect and mass for asymptotically
hyperbolic manifolds from the work [W] we use his definition for
asymptotically hyperbolic manifolds.

\proclaim{Definition 5.1} $(M^n, \ g)$ is said to be an
asymptotically hyperbolic manifold if $(M^n, \ g)$ is a weakly
asymptotically hyperbolic manifold with the standard round sphere
$(S^{n-1}, \ [g_0])$ as its conformal infinity, and, for a geodesic
defining function $\rho$, in the conformally compact coordinates at
the infinity,
$$
g = \sinh^{-2}\rho(d\rho^2 + g_0 + \frac 1n \rho^n h +
O(\rho^{n+1})), \tag 5.1
$$
where $h$ is symmetric two tensors on $S^{n-1}$ at each point.
\endproclaim

In the light of the above definition, we set up a conformally
compact coordinate at the infinity associated with a defining
function $\rho$ as follows. Let
$$
\psi: M\setminus M_c\to R^n\setminus B_{r_0}(0),
$$
for some compact subset $M_c\subset M$, such that
$$
g_H = \frac {(d|x|)^2}{1 + |x|^2} + |x|^2 g_0 =
\sinh^{-2}\rho(d\rho^2 + g_0) \tag 5.2
$$
for $|x| > r_0$ and $\sinh\rho = \frac 1{|x|}$.

We construct an approximate Green's function of an asymptotically
hyperbolic manifold $(M^n, \ g)$. At each point $y\in R^n\setminus
B_{r_0}(0)$, we consider the hyperbolic space $H^n$ in the
coordinate so that
$$
g_H (x) = \frac 1{1 + r_y^2(x)}dr^2 + r^2_y(x)g_0 = (\tilde
g_H)_{ij}(x)dx_idx_j,
$$
where
$$
r_y(x) = \sqrt{A_{ij}(y)x_ix_j}. \tag 5.3
$$
This coordinate can be made into the standard coordinate by the
linear transformation $B: R^n\to R^n$ such that $B^2  = A$. More
importantly we need to ask
$$
(\tilde g_H)_{ij} (y) = g_{ij}(y). \tag 5.4
$$
A simple calculation yields
$$
(\tilde g_H)_{ij}(x) = A_{ij} - \frac {A_{ik}x_kA_{jl}x_l}{1 +
A_{kl}x_kx_l}. \tag 5.5
$$
Hence
$$
A_{ij}(y) = g_{ij}(y) + \frac {g_{ik}(y)y_kg_{jl}(y)y_l}{1 -
g_{kl}(y)y_ky_l}. \tag 5.6
$$
Therefore, since
$$
g_{ij} (x) = \delta_{ij} - \frac {x_ix_j}{1 + |x|^2} +
O(|x|^{-n})\tilde h_{ij}(x) \tag 5.7
$$
and
$$
\tilde h_{ij}(x)x_j = 0, \tag 5.8
$$
we have
$$
A_{ij}(y) = g_{ij}(y) + \frac {y_iy_j}{1+|y|^2} = \delta_{ij} +
O(|y|^{-n})\tilde h_{ij}(y). \tag 5.9
$$
Let $\tilde d_H(x, y)$ be the hyperbolic distance function in the
metric $(\tilde g_H)_{ij}(x)dx_idx_j$ and let
$$
G_y(x) = \frac 1{(n-2)\text{vol}(S^{n-1})}\frac{\theta(\cosh \tilde
d_H(x, y))}{\sinh^{n-2}\tilde d_H(x, y)\cosh^2\tilde d_H(x, y)}.
\tag 5.10
$$
In the geodesic ball $B_1(y)$ in the metric $g$ we calculate
$$
g_{ij} (x) = (\tilde g_H)_{ij} (x)  + \tilde d_H(x, y)O(|y|^{-n})
\tag 5.11
$$
and
$$
\aligned \Delta_g & =\frac 1{\sqrt{\det g}}\partial_j(\sqrt{\det g}
g^{ij}\partial_j)\\ & =  \Delta_H + \tilde d_H(x, y)
O(|y|^{-n})\Delta_H +  (\tilde g_H)^{ij}\partial_i(O(|y|^{-n})
\tilde d_H(x, y))\partial_j.\endaligned \tag 5.12
$$
Thus, for any $x\in B_1(y)$ and $x\neq y$,
$$
\Psi_y(x) = -\Delta_g G_y(x) + nG_y(x) = O(|y|^{-n})O(\tilde d_H(x,
y)^{-n+1}), \tag 5.13
$$
as $|x-y|\to 0$ and $|y|\to \infty$. On the other hand, outside the
geodesic ball $B_1(y)$, we simply need
$$
g_{ij}(x) = (\tilde g_H)_{ij} (x) + O(|x|^{-n})\tilde h_{ij}(x) +
O(|y|^{-n})\xi_{ij}(x,y),
$$
as $|x|\to \infty$ and $|y|\to \infty$, which follows from some
calculations, where
$$
\xi_{ij}(x, y) = \tilde h_{ij} (y) - \frac {\tilde h_{ik}(y)x_kx_j +
\tilde h_{jk}x_kx_i}{1 + |x|^2} + \frac {x_ix_j}{1+|x|^2}\frac
{\tilde h_{kl}x_kx_l}{1+|x|^2}.
$$
Therefore
$$
\xi_{ij}x_j = \frac {\tilde h_{ij}x_j}{1+|x|^2} - \frac
{x_i}{1+|x|^2}\frac {\tilde h_{kl}x_kx_l}{1+|x|^2}
$$
and
$$
(\tilde g_H)^{ij} = \delta_{ij} +x_ix_j + O(|y|^{-n})\xi_{ij}.
$$
This implies
$$
g^{ij} = (\tilde g_H)^{ij} + O(|y|^{-n})\xi_{ij} + O(|x|^{-n})\tilde
h_{ij}+ \ \text{higher order terms}.
$$
Here we use the facts that
$$
(\delta_{ik}+x_ix_k)\xi_{kl}(\delta_{lj} + x_lx_j) = \xi_{ij}
$$
and
$$
(\delta_{ik}+x_ix_k)\tilde h_{kl}(\delta_{lj} + x_lx_j) = \tilde
h_{ij}.
$$
Therefore, outside the geodesic ball $B_1(y)$,
$$
\aligned \Delta_g & = \Delta_{\tilde g_H} + (O(|y|^{-n}) +
O(|x|^{-n}))\Delta_{\tilde g_H}
\\ & \hskip 0.45in + (\tilde g_H)^{ij}\partial_i(O(|y|^{-n}) +
O(|x|^{-n}))\partial_j.\endaligned
$$
One last calculation we need is an estimate for $\Psi_y(x)$ outside
the geodesic ball $B_1(y)$. We compute
$$
\partial_i\tilde d_H(x, y) = \frac 1{\sinh\tilde d_H(x, y)}
(\frac {A_{ik}(y)x_k}{t_x}t_y - A_{ik}(y)y_k),
$$
$$
G'(s) = c_n(- \frac {(n-2)\theta(\cosh s)}{\sinh^{n-1} s \cosh s} -
\frac {2\theta(\cosh s)}{\sinh^{n-3} s\cosh^3 s} + \frac
{\theta'(\cosh s)}{\sinh^{n-3} s\cosh^2 s}),
$$
and
$$
\cosh^n \tilde d_H(y, x) G'(\tilde d_H(y, x)) \to -nc_n,
$$
as $\tilde d_H(y, x)\to \infty$. Thus, outside the geodesic ball
$B_1(y)$,
$$
\Psi_y(x) =- \Delta_g G_y(x) + nG_y(x) =  (O(|x|^{-n}) +
O(|y|^{-n})) O(\frac 1{\cosh^n \tilde d_H (x, y)}). \tag 5.14
$$

\proclaim{Lemma 5.2} Suppose that $(M^n, \ g)$ is an asymptotically
hyperbolic manifold. Then
$$
-\Delta G_y(x) + nG_y(x) = \delta_y(x) + \Psi_y(x) \tag 5.15
$$
where $\Psi_y(x)$ satisfies the estimates (5.13) and (5.14).
\endproclaim

As a consequence we have the following integral representation.

\proclaim{Proposition 5.4} Suppose that $(M^n, \ g)$ is an
asymptotically hyperbolic manifold and that
$$
\psi: M\setminus M_c \to R^n\setminus B_{r_0}(0)
$$
is a conformally compact coordinate associated with a defining
function $\rho$ in which
$$
g = \sinh^{-2}\rho (d\rho^2 + g_0 + \frac {\rho^n}n h +
O(\rho^{n+1})).
$$
Suppose that $v\in C^{2, \alpha}_\delta(M)$ solves the equation
$$
-\Delta v + nv + fv = w\in C^{0, \alpha}_\delta(M),
$$
where $f \in C^{0, \alpha}_\delta(M)$ and $\delta \in (0, n)$. Then,
for each $x\in R^n\setminus B_{r_0}(0)$,
$$
\aligned v(x) & = - \int_{R^n\setminus B_{r_0}(0)}
v(y)\Psi_x(y)dvol_g(y) \\ & \hskip 0.2in + \int_{R^n\setminus
B_{r_0}(0)}(w(y) - f(y)v(y))G_x(y)dvol_g(y) \\ & \hskip 0.2in -
\int_{\partial B_{r_0}(0)}\frac{\partial G_x}{\partial
n}(y)v(y)d\sigma_g(y) \\ & \hskip 0.2in + \int_{\partial
B_{r_0}(0)}\frac{\partial v}{\partial n}(y)G_x(y)d\sigma_g(y).
\endaligned \tag 5.16
$$
\endproclaim

\demo{Proof} We use the density property (cf. [L2]) of the the space
$C^{2, \alpha}_\delta(M)$ to have a sequence of functions $v_n\in
C^\infty_c(M)$ such that
$$
v_n \to v \quad \text{in} \ C^{2, \alpha}_\delta(M).
$$
Then from (5.15) we have, for $v_n$,
$$
\aligned v_n(x) & = - \int_{R^n\setminus B_{r_0}(0)}
v_n(y)\Psi_x(y)dvol_g(y) \\ & \hskip 0.2in + \int_{R^n\setminus
B_{r_0}(0)}(-\Delta v_n + nv_n))G_x(y)dvol_g(y) \\ & \hskip 0.2in -
\int_{\partial B_{r_0}(0)}\frac{\partial G_x}{\partial
n}(y)v_n(y)d\sigma_g(y) \\ & \hskip 0.2in + \int_{\partial
B_{r_0}(0)}\frac{\partial v_n}{\partial n}(y)G_x(y)d\sigma_g(y).
\endaligned \tag 5.17
$$
Hence, by taking the limit, we obtain (5.16) for $v$.
\enddemo

Now we are ready to state and prove our main result of this section.

\proclaim{Theorem 5.5} Suppose that $(M^n, \ g)$ is an
asymptotically hyperbolic manifold and that
$$
\psi: M\setminus M_c \to R^n\setminus B_{r_0}(0)
$$
is a conformally compact coordinate associated with a defining
function $\rho$ in which
$$
g = \sinh^{-2}\rho (d\rho^2 + g_0 + \frac {\rho^n}n h +
O(\rho^{n+1})).
$$
Suppose that $v\in C^{2, \alpha}_\delta(M)$ with $\delta > 0$ solves
the equation
$$
-\Delta v + nv + fv = w
$$
with
$$
f\in C^{0, \alpha}_\kappa(M) \ \text{and} \ w\in C^{2,
\alpha}_\eta(M)
$$
for some $\kappa > 2$ and $\eta>n+1$. Then, for each $x\in
R^n\setminus B_{r_0}(0)$,
$$
v(x) = A(\frac x{|x|})|x|^{-n} + O(|x|^{-(n+1)}). \tag 5.18
$$
\endproclaim

\proclaim{Remark 5.6} We would like to point out that the expansion
(5.18) is a simple consequence of the work in [AC] [MM]. But we need
some explicit expression of the coefficient $A$ in (5.18) to prove
Theorem 6.3 and Lemma 6.5 in the following section, which we did not
find that it is easier to extract it from [AC] [MM] than to obtain
it in the way presented here. The explicit expression of $A$ will be
obtained in the course of the following proof of Theorem 5.5 based
on the integral representation of the solution $v$ in (5.16).
\endproclaim

\demo{Proof of Theorem 5.5} We are going to study the asymptotic
behavior of $v(x)$ term by term in (5.16). We treat the easy ones
first. First we consider
$$
|x|^n\int_{\partial B_{r_0}(0)} \frac{\partial v}{\partial
n}(y)G_x(y)d\sigma(y),
$$
as $|x|\to \infty$ and $y\in \partial B_{r_0}(0)$. Now
$$
|x|^nG_x(y) =\frac {|x|^n} {\cosh^n \tilde d_H(y, x)} \frac
{c_n\cosh^{n-2}\tilde d_H(y, x)}{\sinh^{n-2}\tilde d_H(y,
x)}\theta(\cosh \tilde d_H(y, x)),
$$
where
$$
\cosh \tilde d_H(y, x) = t_xt_y - A_{ij}(x)x^i \ y^j
$$
and
$$
A_{ij}(x) =\delta_{ij} + O(|x|^{-n}).
$$
Hence
$$
|x|^nG_x(y) =\frac 1 {(\frac {t_x}{|x|}t_y - A_{ij} (x) \frac
{x^i}{|x|} \ y^j)^n} \frac {c_n\cosh^{n-2}\tilde d_H(y,
x)}{\sinh^{n-2}\tilde d_H(y, x)}\theta(\cosh \tilde d_H(y, x))\in
C^1(M)
$$
and
$$
\lim_{\lambda\to\infty} \lambda^nG_{\lambda\frac x{|x|}} (y) =c_n
(t_y - \frac x{|x|}\cdot y)^{-n}.
$$
Therefore
$$
|x|^n \int_{\partial B_{r_0}(0)} \frac{\partial v}{\partial
n}(y)G_x(y)d\sigma(y)\in C^1(M)
$$
and
$$
\aligned A_1(\frac x{|x|}) & = \lim_{\lambda\to\infty}
\lambda^n\int_{\partial B_{r_0}(0)} \frac{\partial v}{\partial
n}(y)G_{\lambda\frac x{|x|}}(y)d\sigma(y) \\ & = c_n\int_{\partial
B_{r_0}(0)} \frac{\partial v}{\partial n}(y)(t_y - \frac x{|x|}\cdot
y)^{-n}d\sigma(y). \endaligned\tag 5.19
$$
Next we consider
$$
|x|^n \int_{\partial B_{r_0}(0)}\frac{\partial G_x}{\partial
n}(y)v(y)d\sigma(y),
$$
as $|x|\to \infty$ and $y\in \partial B_{r_0}(0)$. We compute
$$
\aligned |x|^n & \frac{\partial G_x}{\partial n}(y) =
|x|^n\rho(y)c_nG'(\tilde d_H(y, x)) \frac {\partial \tilde d_H(y, x)}{\partial r} \\
& = |x|^n\rho(y)G' \frac {t_x\frac {g_{ij}y^i \ y^j}{|y|t_y} -
g_{ij}x^i \ \frac {y^j}{|y|}}{\sinh \tilde d_H(y, x)},\endaligned
$$
where
$$
G'(s) = c_n(- \frac {(n-2)\theta(\cosh s)}{\sinh^{n-1} s \cosh s} -
\frac {2\theta(\cosh s)}{\sinh^{n-3} s\cosh^3 s} + \frac
{\theta'(\cosh s)}{\sinh^{n-3} s\cosh^2 s})
$$
and
$$
\cosh^n \tilde d_H(y, x) G'(\tilde d_H(y, x)) \to -nc_n
$$
as $\tilde d_H(y, x)\to \infty$. Therefore
$$
\aligned A_2 & (\frac x{|x|})= \lim_{\lambda\to\infty}
\lambda^n\int_{\partial B_{r_0}(0)}\frac{\partial G_{\lambda\frac
x{|x|}}}{\partial n}(y)v(y)d\sigma(y) \\ & = -nc_n\int_{\partial
B_{r_0}(0)}v(y)(t_y - \frac x{|x|}\cdot y)^{-n}\frac {\frac
{|y|}{t_y} - \frac {x\cdot y}{|x||y|}}{t_y - \frac x{|x|}\cdot
y}d\sigma(y).
\endaligned\tag 5.20
$$
For the term
$$
|x|^n\int_{R^n\setminus B_{r_0}(0)} (h-fv)G_x(y)dvol_g(y),
$$
we know, for any given $y\in R^n\setminus B_{r_0}(0)$,
$$
\lim_{\lambda\to\infty} \lambda^nG_{\lambda\frac x{|x|}} (y) =c_n
(t_y - \frac x{|x|}\cdot y)^{-n}.
$$
We observe that
$$
t_y - \frac x{|x|}\cdot y = t_y - |y|\cos \phi\geq (1-\cos\phi)|y|,
$$
where $\phi$ is the angle between $x$ and $y$. Fixing a direction
$\frac x{|x|}$, we easily see that for any $\epsilon_0 > 0$
$$
\aligned \lim_{\lambda\to\infty}& \lambda^n\int_{\{y\in R^n\setminus
B_{r_0}(0): \ \cos\phi \leq 1-\epsilon_0\}} (h-fv)G_{\lambda \frac
x{|x|}}(y)dvol_g(y) \\ & = \int_{\{y\in R^n\setminus B_{r_0}(0): \
\cos\phi \leq 1-\epsilon_0\}} (h-fv)(t_y - \frac x{|x|}\cdot
y)^{-n}dvol_g(y).\endaligned
$$
On the other hand, when $\cos\phi> 1- \epsilon_0$, it suffices to
verify the claim
$$
\int_{r_0}^\infty\int_{\{\cos\phi > 1-\epsilon_0\}} (h - fv) (t_y-
r\cos\phi)^{-n} \frac {r^{n-1}}{t_y}d\sigma_0dr < \infty. \tag 5.21
$$
Here we need to use the fact that $\eta > n$. We simply notice that
$$
t_y - |y|\cos\phi = \frac {1 + \sin^2\phi|y|^2}{t_y + |y|\cos\phi}.
$$
Hence
$$
\aligned \int_{\{\cos\phi > 1-\epsilon_0\}} &  (t_y -
|y|\cos\phi)^{-n}d\sigma \lessapprox
\int_0^{\epsilon_0}\int_{S^{n-2}} (t_y -
|y|\cos\phi)^{-n}\phi^{n-2}d\sigma d\phi \\ & \lessapprox
\int_0^{\epsilon_1}\int_{S^{n-2}} (t_y -
|y|\cos\phi)^{-n}\phi^{n-2}d\sigma d\phi  \\ & \hskip 0.5in +
\int_{\epsilon_1}^{\epsilon_0}\int_{S^{n-2}} (t_y -
|y|\cos\phi)^{-n}\phi^{n-2}d\sigma d\phi \\ & \lessapprox
|y|^{n}\epsilon_1^{n-1} + |y|^{-n}\epsilon_1^{-n-1} \lessapprox |y|
\endaligned
$$
for $\epsilon_1 = |y|^{-1} < \epsilon_0$. Therefore
$$
\int_{\{\cos\phi > 1-\epsilon_0\}} (h - fv) (t_y- r\cos\phi)^{-n}
\frac {r^{n-1}}{t_y}d\sigma_0 = O(r^{-\iota + n -1}),
$$
where $\iota = \min\{\eta, n+\frac 12 \delta\} > n$, which implies
our claim (5.21). Thus
$$
\aligned A_0(\frac x{|x|}) = \lim_{\lambda\to\infty}&
\lambda^n\int_{R^n\setminus B_{r_0}(0)} (h-fv)G_{\lambda \frac
x{|x|}}(y)dvol_g(y) \\ & = \int_{R^n\setminus B_{r_0}(0)} (h-fv)(t_y
- \frac x{|x|}\cdot y)^{-n}dvol_g(y).\endaligned \tag 5.22
$$
A similar argument yields the next order when we have $\kappa > 2$
and $\eta > n+1$. For the last term
$$
|x|^n\int_{R^n\setminus B_{r_0}(0)} v(y)\Psi_x(y)dvol_g(y),
$$
we need to use the estimates about the correction term $\Psi_x(y)$
in (5.13) and (5.14). We first look at
$$
\aligned |x|^n & \int_{B_1(x)}v(y)\Psi_x(y)dvol_g(y) \lessapprox
|x|^n\int_0^1\int_{S^{n-1}}v(y)\Psi_x(y)\sinh^{n-1}r d\sigma dr \\
& \lessapprox |x|^n\int_0^1
|y|^{-n+\epsilon}|x|^{-n}r^{-n+1}r^{n-1}dr\endaligned
$$
for any small positive number $\epsilon$. Clearly
$$
\lim_{|x|\to\infty}|x|^n \int_{B_1(x)}v(y)\Psi_x(y)dvol_g(y) = 0
\tag 5.23
$$
since $|y| \geq c |x|$ for $y\in B_1(x)$ and $|x|\to \infty$. Next
we look at
$$
|x|^n \int_{(R^n\setminus B_{r_0}(0))\setminus
B_1(x)}v(y)\Psi_x(y)dvol_g(y).
$$
In the light of (5.14) and (5.23), using the argument we used to
treat last term to obtain (5.21) and (5.22), we have
$$
\aligned A_{-1}(\frac x{|x|}) & = \lim_{\lambda\to\infty}\lambda^n
\int_{R^n\setminus B_{r_0}(0)}v(y)\Psi_{\lambda \frac
x{|x|}}(y)dvol_g(y) \\ & =\lim_{\lambda\to\infty}\lambda^n
\int_{(R^n\setminus B_{r_0}(0))\setminus B_1(x))}v(y)\Psi_{\lambda
\frac x{|x|}}(y)dvol_g(y).\endaligned \tag 5.24
$$
We have thus proven the theorem with
$$
A(\frac x{|x|}) =A_{-1}(\frac x{|x|}) + A_0(\frac x{|x|}) +
A_1(\frac x{|x|}) + A_2(\frac x{|x|}). \tag 5.25
$$
\enddemo

\head 6. Proof of the main theorem\endhead

In this section we prove the main theorem. We first recall a
positive mass theorem for asymptotically hyperbolic manifolds from
[W]. Readers are referred to [CH] for more elaborated and complete
discussions of positive mass theorems for asymptotically hyperbolic
manifolds. Recall that, on an asymptotically hyperbolic manifold
$(M^n, \ g)$ as defined in Definition 5.1, we have a coordinate at
the infinity such that
$$
g = \sinh^{-2}\rho (d\rho^2 + g_0 + \frac {\rho^n}n h +
O(\rho^{n+1})). \tag 6.1
$$
In [W] it was proven that

\proclaim{Theorem} (Xiaodong Wang) Suppose that $(M^n, \ g)$ is a
spin asymptotically hyperbolic manifold and that $R_g \geq -n(n-1)$.
Then
$$
\int_{S^{n-1}} \text{Tr}_{g_0}h (x)dvol_{g_0}(x) \geq
|\int_{S^{n-1}}\text{Tr}_{g_0}h (x)xdvol_{g_0}(x)|. \tag 6.2
$$
Moreover the equality holds if and only if $(M^n, \ g)$ is isometric
to the standard hyperbolic space $H^n$.
\endproclaim

We adopt the idea from [M] to deal with asymptotically hyperbolic
manifolds with corners along a hypersurface.

\proclaim{Definition 6.1} A Riemannian manifold $(M^n, \ g)$ is said
to have corners along a hypersurface $\Sigma$ if there is a smooth
embedded hypersurface $\Sigma\subset M$ such that $M\setminus \Sigma
= M_-\bigcup M_+$ and the inside $(M_-, \ g_-)= (M_-, \ g)$ is a
smooth compact Riemannian manifold with a boundary $\Sigma$ and the
outside $(M_+, \ g_+)= (M_+, \ g)$ is a smooth Riemannian manifold
with a boundary $\Sigma$. Moreover $g_-$ and $g_+$ agree on the
boundary $\Sigma$, that is, $g$ continuous across the hypersurface
$\Sigma\subset M$.
\endproclaim

We will consider the outward mean curvature $H_-$ of the
hypersurface $\Sigma$ in $(M_-, \ g_-)$ and the inward mean
curvature $H_+$ of the hypersurface $\Sigma$ in $(M_+, \ g_+)$. Near
the hypersurface $\Sigma$ we may use Gauss coordinates, that is, for
some $\nu_0>0$, a point $p$ within distance $\nu_0$ from the
hypersurface $\Sigma$ is labeled by a point $x$ on the hypersurface
$\Sigma$ and the signed distance $d = \text{dist}(p, \Sigma)$ to the
hypersurface $\Sigma$. We now recall the smoothing operation given
in Proposition 3.1 in [M] to have $C^2$ metrics on $M$ approximating
$g$.

\proclaim{Proposition 6.2} (Pengzi Miao) Suppose that $(M, \ g)$ is
a manifold with corners along a hypersurface $\Sigma$. Then there is
a family of $C^2$ metrics $g_\nu$, for $\nu\in (0, \nu_0)$, on $M$
such that $g_\nu$ uniformly converges to $g$ on $M$ and $g_\nu = g$
outside $\Sigma\times (-\frac 12 \nu, \frac 12 \nu)$. Furthermore,
the scalar curvature $R_\nu$ of the metric $g_\nu$ satisfies
$$
\left\{\aligned R_\nu(p) = O(1) \hskip 1.6in & \quad \text{when
$d\in (\frac
{\nu^2}{100}, \frac \nu 2]$}\\
R_\nu (p) = O(1) + 2(H_- - H_+)(\frac {100}{\nu^2}\phi(\frac
{100}{\nu^2})) & \quad \text{when $d \leq \frac {\nu^2}{100}$},
\endaligned\right. \tag 6.3
$$
where $O(1)$ stands for terms bounded independent of $\nu$ and
$\phi(t)\in C^\infty_c(-1,1)$ is a standard mollifier.
\endproclaim

Our next goal is to conformally deform the metric $g_\nu$ so that
the scalar curvature is greater than or equal to $-n(n-1)$ so that
the positive mass theorem in [W] applies. The reason that $g_\nu$
admits such conformal deformation relies on the fact that
$$
\int_M [\frac {n-2}{4(n-1)}(R_\nu+n(n-1))^-]^\frac n2 dvol_{g_\nu}
\leq \epsilon_0^\frac n2
$$
whenever $\nu$ is sufficiently small and $H_- - H_+ \geq 0$. Thus we
are ready to state and prove our main theorem.

\proclaim{Theorem 6.3} Suppose that $(M, \ g)$ is a spin Riemannian
manifold with corners along a hypersurface $\Sigma$ and that the
outside is an asymptotically hyperbolic manifold and the inside is
compact. Suppose that the scalar curvature of both the inside and
outside metrics are greater than or equal to $-n(n-1)$ and that
$$
H_- (x) \geq H_+ (x)
$$
for each $x$ on the hypersurface. Then, if in a coordinate system at
the infinity,
$$
g = \sinh^{-2}\rho (d\rho^2 + g_0 + \frac {\rho^n}n h +
O(\rho^{n+1})),
$$
then
$$
\int_{S^{n-1}} \text{Tr}_{g_0}h (x)dvol_{g_0}(x) \geq
|\int_{S^{n-1}}\text{Tr}_{g_0}h (x)xdvol_{g_0}(x)|.
$$
\endproclaim

\demo{Proof} We first use the smoothing operation given in [M] as
stated in the above proposition. For each small $\nu < \nu_0$, we
then solve the equation
$$
-\Delta_{g_\nu} v + nv + f_\nu v = - f_\nu \tag 6.4
$$
on $M$ for
$$
f_\nu = - \frac {n-2}{4(n-1)}(R_\nu + n(n-1))^-.
$$
According to Proposition 6.2 above
$$
\int_M f_\nu^\frac n2 dvol_{g_\nu} \leq C(g)\nu,
$$
where $C(g)$ depends only on the metric $g$. For sufficiently small
$\nu$ we apply Proposition 3.2 in Section 3 to obtain a positive
solution $v_\nu$ to the above equation (6.4). Then we consider the
new metric
$$
\tilde g_\nu = (1+v_\nu)^\frac 4{n-2}g_\nu.
$$
In the light of Lemma 3.1 in Section 3 we know that the scalar
curvature $\tilde R_\nu$ of the new metric $\tilde g_\nu$ is greater
than or equal to $-n(n-1)$. To finish the proof we need to establish
the following two lemmas.
\enddemo

\proclaim{Lemma 6.4} Suppose that $(M^n, \ g)$ is an asymptotically
hyperbolic manifold and in a coordinate at the infinity associated
with a geodesic defining function $r$
$$
g = \sinh^{-2}\rho(d\rho^2 + g_0 + \frac {\rho^n}n h +
O(\rho^{n+1})),
$$
where
$$
r = \frac {\cosh\rho -1}{\sinh\rho}.
$$
And suppose that
$$
v = A(\frac {x}{|x|})\rho^n + O(\rho^{n+1})
$$
is a positive function on $M$. Then there is a geodesic defining
function $\tilde r$ for $\tilde g = (1+v)^\frac 4{n-2}g$ such that
$$
\tilde g = \sinh^{-2}\tilde\rho(d\tilde\rho^2 + g_0 + \frac
{\tilde\rho^n}n \tilde h + O(\tilde\rho^{n+1})),
$$
where
$$
\tilde r = \frac {\cosh\tilde\rho -1}{\sinh\tilde\rho}
$$
and
$$
\tilde h = \frac {4(n+1)}{n-2}A(\frac x{|x|})g_0 + h. \tag 6.5
$$
\endproclaim

\demo{Proof} First we recall that the geodesic defining function of
the metric $g$ is a defining function $s$ such that
$$
|ds|_{s^2g} = 1
$$
near the infinity. We refer the readers to Lemma 2.1 in [G] for the
existence and uniqueness of the geodesic defining function
associated with each boundary metric in the conformal infinity. We
start with a geodesic defining function $r$ for $g$. Then for each
$\theta\in S^{n-1}$, let
$$
\tilde r = e^w r \quad \text{and} \quad w(\theta, 0)  =0.
$$
By the definition, $w$ satisfies
$$
2\frac{\partial w}{\partial r} + r|dw|^2_{r^2g} = \frac
1r((1+v)^\frac 4{n-2} - 1) = \frac 4{n-2}Ar^{n-1} + O(r^n). \tag 6.6
$$
By an inductive argument we obtain
$$
\frac {\partial^k w}{\partial r^k}(\theta, 0) = 0
$$
for $k\leq n-1$ and
$$
\frac {\partial^n w}{\partial r^n}(\theta, 0) = (n-1)!\frac 2{n-2}
A(\theta). \tag 6.7
$$
Hence
$$
w(\theta, r) = \frac 2{n(n-2)}A(\theta)r^n + O(r^{n+1}).
$$
This gives
$$
\tilde r (\theta, r) = r + \frac 2{n(n-2)}A(\theta)r^{n+1} +
O(r^{n+2}). \tag 6.8
$$
By the construction of the coordinate associated with a geodesic
defining function, we need to compare the integral curves of the
vector field $\frac {\partial}{\partial r}$ and $\frac {\partial
}{\partial \tilde r}$. We know
$$
d\tilde r = (1 + \frac {2(n+1)}{n(n-2)}Ar^n )dr + \frac
2{n(n-2)}r^{n+1}\frac {\partial A}{\partial \theta_i} d\theta_i +
O(r^{n+1}),
$$
which implies
$$
\aligned \frac {\partial}{\partial \tilde r} & = (1+v)^{-\frac
4{n-2}}(1 + \frac {2(n+1)}{n(n-2)}Ar^n) \frac {\partial}{\partial r}
\\ & \hskip 0.3in + (1+v)^{-\frac 4{n-2}}(r^{n+1}\frac
2{n(n-2)}\frac {\partial A}{\partial \theta_j} +
O(r^{n+2}))g_r^{ij}\frac {\partial }{\partial \theta_i} \\ & = \frac
{\partial}{\partial r} - \frac {2(n-1)}{n(n-2)}Ar^n \frac
{\partial}{\partial r} + O(r^{n+1})\endaligned \tag 6.9
$$
Therefore
$$
\tilde\theta(\theta, r) =  \theta + O(r^{n+1}). \tag 6.10
$$
Thus
$$
\sinh^2\tilde\rho \ \tilde
g(\frac{\partial}{\partial\tilde\theta_i},
\frac{\partial}{\partial\tilde\theta_j})= \sinh^2\tilde\rho \
(1+v)^\frac 4{n-2} g(\frac{\partial}{\partial\theta_i},
\frac{\partial}{\partial\theta_j}) + O(r^{n+1}).
$$
In the light of the fact that
$$
\aligned  \frac {\sinh\tilde\rho}{1+\cosh\tilde\rho} & = \tilde r =
r(1 + \frac 2{n(n-2)}Ar^n+ O(r^{n+1})) \\ &  = \frac {\sinh\rho}{1 +
\cosh\rho}(1 + \frac 2{n(n-2)}Ar^n+ O(r^{n+1}))\endaligned
$$
we have
$$
\sinh^2\tilde\rho = \sinh^2\rho (\frac {1 + \cosh\tilde\rho}{1 +
\cosh\rho })^2 (1 + \frac 4{n(n-2)}Ar^n+ O(r^{n+1})),
$$
where
$$
\aligned \frac {1 + \cosh\tilde\rho}{1 + \cosh\rho } & = 1 + \frac
{\cosh\tilde\rho-\cosh\rho}{1 + \cosh\rho }
\\ & = 1 + O(r)(\tilde\rho - \rho) \\ & = 1 + O(r)(\tanh^{-1} \tilde r -
\tanh^{-1} r) \\ & = 1 + O(r^{n+1}).\endaligned
$$
Finally, we arrive at
$$
g_0 + \frac {\tilde\rho^n}n \tilde h + O(\tilde\rho^{n+1}) = g_0 +
(\frac {4(n+1)}{n(n-2)} \rho^n A(\theta) g_0 + \frac {\rho^n}n
h(\theta)) + O(\rho^{n+1}), \tag 6.11
$$
which gives
$$
\tilde h = \frac{4(n+1)}{n-2} A(\frac x{|x|})g_0 + h. \tag 6.12
$$
So the calculation is completed.
\enddemo

The next lemma is an estimate of the perturbation of mass aspect
$\frac {4(n+1)}{n-2}A_\nu(\frac x{|x|})g_0$ in terms of the small
number $\nu$ as $\nu\to 0$ when $v = v_\nu$.

\proclaim{Lemma 6.5} Suppose that $(M, \ g)$ is a complete
Riemannian manifold with corners along a hypersurface and that the
outside is an asymptotically hyperbolic manifold. Suppose that the
scalar curvature of both the inside and outside metrics are greater
than or equal to $-n(n-1)$ and that
$$
H_- (x) \geq H_+ (x)
$$
for each $x$ on the hypersurface. Let $g_\nu$ be constructed as in
Proposition 6.2. Then there is a unique positive solution $v_\nu\in
C^{2,\alpha}_\delta(M)$ to the equation
$$
-\Delta_{g_\nu} v + nv - \frac {n-2}{4(n-1)}(R_\nu+ n(n-1))^-v =
\frac {n-2}{4(n-1)}(R_\nu + n(n-1))^-.
$$
when $\nu$ is sufficiently small. Moreover, in a coordinate at the
infinity associated with a geodesic defining function $r$,
$$
v_\nu = A_\nu(\frac x{|x|})r^n + O(r^{n+1})
$$
and
$$
|A_\nu(\frac x{|x|})| \leq C \nu^\frac 1{n+1}, \tag 6.13
$$
where $C$ is independent of $\nu$.
\endproclaim

\demo{Proof} By Proposition 6.2 we have
$$
\frac {n-2}{4(n-1)}(R_\nu + n(n-1))^- \leq C
$$
with compact support inside $\partial\Omega\times [-\frac \nu 2,
\frac \nu 2]$, where $C$ is independent of $\nu$. Hence
$$
\int_M (\frac {n-2}{4(n-1)}(R_\nu+n(n-1))^-)^\frac n2dvol_{g_\nu}
\leq C\nu.
$$
Therefore, by Proposition 3.2 and Theorem 5.5, there is exists the
unique positive solution to the equation
$$
-\Delta_{g_\nu} v + nv - \frac {n-2}{4(n-1)}(R_\nu+ n(n-1))^-v =
\frac {n-2}{4(n-1)}(R_\nu + n(n-1))^-.
$$
when $\nu$ is sufficiently small and in a coordinate at the infinity
associated with a geodesic defining function $r$,
$$
v_\nu = A_\nu(\frac x{|x|})r^n + O(r^{n+1}),
$$
where $A(\frac x{|x|})$ is given in (5.25).

First of all, since
$$
\|\frac {n-2}{4(n-1)}(R_\nu+n(n-1))^-\|_{W^{0, n+1}_\gamma(M)} \leq
C \nu^\frac 1{n+1} \tag 6.14
$$
for any $\gamma$, we know by an isomorphism theorem similar to
Proposition 2.3 (cf. Theorem C in [L2]), that
$$
\|v_\nu\|_{W^{2, n+1}_\gamma(M)} \leq C \nu^\frac 1{n+1}
$$
for any $\gamma < \frac {n+1}2$. Then by the Sobolev embedding
Theorem ([L2]) we have
$$
\|v_\nu\|_{C^{1, \alpha}_\gamma(M)} \leq C\nu^\frac 1{n+1} \tag 6.15
$$
for some $\alpha\in (0, 1)$.

Next we estimate $A(\frac x{|x|})$ term by term. We treat the easy
terms first. For the term
$$
A_0(\frac x{|x|}) = c_n\int_{R^n\setminus
B_{r_0}(0)}(R_\nu+n(n-1))^-(1+v)(\sqrt{1+ |y|^2} - \frac x{|x|}\cdot
y)^{-n}dvol_{g_\nu}(y),
$$
we simply ask $r_0$ is large enough so the support of
$(R_\nu+n(n-1))^-$ is outside of $R^n\setminus B_{r_0}(0)$.
Therefore, we may choose $r_0$ so that
$$
A_0(\frac x{|x|}) = 0. \tag 6.16
$$
For the term
$$
A_1(\frac x{|x|}) = c_n\int_{\partial B_{r_0}(0)}\frac {\partial
v_\nu}{\partial n}(\sqrt{1+|y|^2} - \frac x{|x|}\cdot
y)^{-n}d\sigma_{g_\nu}(y),
$$
we easily see that
$$
A_1(\frac x{|x|}) \leq C \nu^\frac 1{n+1}. \tag 6.17
$$
Similarly, for the term
$$
A_2(\frac x{|x|}) =-nc_n \int_{\partial B_{r_0}(0)}v(y)
(\sqrt{1+|y|^2} - \frac x{|x|}\cdot y)^{-n}\frac {\frac
{|y|}{\sqrt{1+|y|^2}} - \frac x{|x|}\cdot \frac
y{|y|}}{\sqrt{1+|y|^2} - \frac x{|x|}\cdot y}d\sigma_{g_\nu}(y),
$$
we easily derive from (6.15) that
$$
|A_2(\frac x{|x|})| \leq C \nu^{\frac 1{n+1}}.\tag 6.18
$$
The last term is
$$
A_{-1}(\frac x{|x|}) = \lim_{\lambda\to\infty}\lambda^n
\int_{R^n\setminus B_{r_0}(0)} v_\nu (y) \Psi_{\lambda \frac
x{|x|}}(y)dvol_g(y).
$$
Due to (6.15) and the estimate (5.14) we know
$$
\aligned |A_{-1}| & \leq C \nu^\frac
1{n+1}\lim_{\lambda\to\infty}\lambda^n \int |y|^{-\frac n2}
O(|y|^{-n})O(\frac 1{\cosh^n d_H(\lambda\frac x{|x|},
y)})dvol_g(y)\\ & \leq C \nu^\frac 1{n+1},\endaligned \tag 6.19
$$
where in the last step we use the same argument we used to establish
(5.21) and (5.22) to deal with the term $(\sqrt{1+|y|^2} - \frac
x{|x|}\cdot y)^{-n}$, which is only big when $\frac y{|y|}$ is very
close to $\frac x{|x|}$. Thus we have proved that
$$
|A_\nu(\frac x{|x|})| \leq C\nu^\frac 1{n+1} \tag 6.20
$$
for some $C$ independent of $\nu$.
\enddemo

\demo{Proof of Theorem 6.3} To finish the proof of Theorem 6.3 we
simply notice that for each $\nu$ sufficiently small, by Lemma 3.1
in Section 3, we may apply the positive mass theorem in [W] to the
metric $(1+v_\nu)^\frac 4{n-2}g_\nu$ and obtain that
$$
\int_{S^{n-1}} \text{Tr}_{g_0}\tilde h dvol_{g_0}(x)\geq
|\int_{S^{n-1}}\text{Tr}_{g_0}\tilde h xdvol_{g_0}(x)|
$$
where
$$
\tilde h = \frac {4(n+1)}{n-2}A_\nu(\frac x{|x|}) g_0 + h.
$$
Here we note that the mass aspect of $g_\nu$ is the same as the mass
aspect of $g$ since $g_\nu$ is the same as $g$ outside a compact
set. Therefore, as $\nu\to 0$, we have
$$
\int_{S^{n-1}}\text{Tr}_{g_0}h dvol_{g_0}(x) \geq
|\int_{S^{n-1}}\text{Tr}_{g_0}h x dvol_{g_0}(x)|.
$$
So the proof is finished.
\enddemo

\vskip 0.2in \noindent {\bf References}:

\roster

\vskip 0.1in
\item"{[A]}" L. Andersson, Elliptic Systems on Manifolds with
Asymptotically Negative Curvature, Indiana Math. Journal, 42(4):
1993, 1359-1388.


\vskip 0.1in
\item"{[AC]}" L. Andersson and P. Chrusciel, Solutions of the constraint
equations in general relativity satisfying "hyperboloidal boundary
conditions". Dissertationes Math. (Rozprawy Mat.) 355 (1996), 100.

\vskip 0.1in
\item"{[AD]}" L. Andersson and M. Dahl, Scalar curvature rigidity for
asymptotically hyperbolic manifolds, Ann. Global Anal. and Geo.
16(1998) 1-27.

\vskip 0.1in
\item"{[B]}" R. Bartnik, New definition of quasi-local mass, Phys.
Rev. Lett. 62 (20) 1989, 2346-2348.

\vskip 0.1in
\item"{[BMQ]}" V. Bonini, P. Miao and J. Qing, Ricci curvature
rigidity for weakly asymptotically hyperbolic manifolds,  Comm.
Anal. Geom. 14 (2006), no. 3, 603--612.  ArXiv: hep-th/0310378.

\vskip 0.1in
\item"{[CH]}" P.T. Chru\'{s}ciel and M. Herzlich, The mass of
asymptotically hyperbolic Riemannian manifolds, Pacific J. Math. 212
(2003), no. 2, 231--264. ArXiv: math.DG/0110035.


\vskip 0.1in
\item"{[G]}" C. R. Graham, Volume and Area renormalizations for
conformally compact Einstein metrics. The Proceedings of the 19th
Winter School "Geometry and Physics" (Srn\`{i}, 1999). Rend. Circ.
Mat. Palermo (2) Suppl. No. 63 (2000), 31--42.

\vskip 0.1in
\item"{[GL]}" C. R. Graham and J. Lee, Einstein
Metrics with Prescribed Conformal Infinity on the Ball. Adv. Math.,
87(2):186--225, 1991.

\vskip 0.1in
\item"{[L1]}" J. Lee, The spectrum of an asymptotically hyperbolic
Einstein manifold. Comm. Anal. Geom. 3 (1995), no. 1-2, 253--271.

\vskip 0.1in
\item"{[L2]}" J. Lee, Fredholm Operators and {E}instein Metrics on Conformally Compact
Manifolds. Mem. Amer. Math. Soc. 183 (2006), no. 864, vi+83 pp.
ArXiv: math.DG/0105046.

\vskip 0.1in \item"{[Mz]}" R. Mazzeo,  The Hodge Cohomology of a
Conformally Compact Metric,  J. Differential Geom., 28(2) 1988,
309--339.

\vskip 0.1in
\item"{[MM]}" R. Mazzeo and R. Melrose, Meromorphic
extension of the resolvent on complete spaces with asymptotically
constant negative curvature. J. Funct. Anal. 75 (1987), no. 2,
260--310.

\vskip 0.1in
\item"{[M]}" P. Miao, Positive mass theorem on manifolds
admitting corners along a hypersurface, Adv. Theor. Math. Phys.,
6(6):1163--1182, 2002.

\vskip0.1in
\item"{[Mo]}" M. Min-Oo, Scalar curvature rigidity of
asymptotically hyperbolic spin manifolds, Math. Ann. 285 (1989),
527–539.


\vskip 0.1in
\item"{[PW]}" M. Protter and H. Weinberger, ``The Maximum Principles in differential equations'',
Springer, New York, 1984.

\item"{[Q]}" J. Qing, On the rigidity for conformally
compact Einstein manifolds,  International Mathematics Research
Notices, 21 (2003) 1141-1153, ArXiv: math.DG/0305084.

\vskip .1in
\item"{[R]}" J. Ratcliff, ``Foundations of hyperbolic manifolds'',
GTM 149, Springer Verlag, 1994.

\vskip 0.1in
\item"{[SY]}" R. Schoen and S-T. Yau, On the Proof of
the Positive Mass Conjecture in General Relativity, Comm. Math.
Phys., 65(1):45-76, 1979.


\vskip 0.1in
\item"{[ST1]}" Y. Shi and L. Tam, Positive mass theorem and the
boundary behaviors of a compact manifolds with nonnegative scalar
curvature, J. Differential Geom. 62 (2002), no. 1, 79--125. ArXiv:
math.DG/0301047.

\vskip 0.1in
\item"{[ST2]}" Y. Shi and L. Tam, Boundary behaviors and scalar curvature of
compact manifolds, ArXiv: math.DG/0611253.



\vskip 0.1in
\item"{[W]}" X. Wang, The mass of Asymptotically hyperbolic
manifolds, J. Diff. Geo. 57 (2001), no. 2, 273-299.
\endroster

\enddocument